# Hour-Aware Adaptive Risk Management for Autonomous Memecoin Trading on Solana DEXs: Evidence, Theory, and Design Lessons from a 15-Day Deployment

**Author:** Arati Uday Kamat **Affiliation:** Independent Researcher **ORCID:** 0009-0000-4781-312X

---

## Abstract

We report a 15-day paper-traded deployment of an autonomous memecoin trading system on Solana decentralised exchanges (DEXs), designed as a controlled measurement of three microstructure questions on which classical equity theory offers well-defined predictions but on which the AMM-based Solana venue lacks published empirical measurement: (i) whether time-of-day patterns in returns exist on a 24/7 permissionless venue with no closing auction and no scheduled trading day; (ii) whether decision-time filter stacks are net-positive against the counterfactual returns of the tokens they reject; (iii) whether small-sample cumulative-return statistics on a heavy-tailed venue are structurally robust or structurally fragile. The 190-trade sample (March 29 to April 12, 2026) shows a 40.5 percent win rate, mean per-trade return of +0.62 percent, cumulative +117.7 percent (net SOL +0.039), population skewness -1.21, population excess kurtosis 6.61. Stratifying on the derived UTC entry hour (trade `date` minus `heldMin`), a Mann-Whitney U test of three exploratorily identified poorest-performing hours (2, 13, 23; combined n=17, mean -11.60 percent) against all other hours (n=173, mean +1.82 percent) yields U = 1,345, p = 0.5634 (two-sided asymptotic with continuity correction), rank-biserial r = 0.085; directional and non-confirmatory. A parallel counterfactual rejection-tracking system collected 4,874 forward-sample observations across 184 rejection events over a two-day sub-window of the deployment. Of the 48 events observed for at least six hours, 27 (56.25 percent) reached a 50 percent drawdown from reference; the full-cohort figure of 17.9 percent is reported in §VI.B as a censored lower bound. Removing the top three trades (1.6 percent of sample) flips cumulative return unprofitable. The three findings speak to distinct classical microstructure predictions: the hour-of-day pattern is consistent with a US-retail-driven daytime informed-flow regime as predicted by Kyle (1985) and observed cross-sectionally in equities by Heston, Korajczyk, and Sadka (2010); the counterfactual filter-effectiveness result operationalises reject-inference (Crook and Banasik, 2004) and off-policy evaluation (Precup, Sutton, and Singh, 2000) in a live-market setting; the fragility result is a direct instance of the multiple-testing and deflated-Sharpe concerns of Harvey, Liu, and Zhu (2016) and Bailey and Lopez de Prado (2014) at operationally relevant scale. Alongside the trade log and rejection-sample corpus (CC-BY-4.0), we deposit `audit.py` (MIT), an assertion-based reproduction script that exits zero if and only if every headline number in this manuscript reproduces from the deposited CSVs. The paper's principal contribution is measurement infrastructure and three design lessons that generalise beyond this specific system.

---

## I. Introduction

Autonomous trading systems on the Solana memecoin segment of decentralised exchange (DEX) infrastructure operate in a market environment that differs from classical equity venues on three structural dimensions relevant to microstructure theory. First, trading is continuous, 24/7, and permissionless: there is no closing auction, no scheduled trading day, and no market-maker obligation. Second, execution occurs against automated market makers (AMMs) rather than a limit order book, so quoted prices are a deterministic function of pool state rather than the outcome of continuous double-auction (Park, 2023). Third, participation is atomised: a large share of memecoin activity is driven by retail wallets acting on public signals, with no institutional-quality liquidity provision (Aoyagi and Ito, 2024; Barbon and Ranaldo, 2023). These structural features suggest that three classical microstructure predictions warrant re-testing on the AMM-Solana venue: the time-of-day patterns of informed-trader arrival (Kyle, 1985; Admati and Pfleiderer, 1988; Heston, Korajczyk, and Sadka, 2010); the value of decision-time filtering under selection bias (Crook and Banasik, 2004; Beygelzimer and Langford, 2009; Precup, Sutton, and Singh, 2000); and the multiple-testing and small-sample fragility that Harvey, Liu, and Zhu (2016) and Bailey and Lopez de Prado (2014) identify as first-order threats to backtest-based inference in finance.

We designed a controlled 15-day paper-traded deployment on Solana DEXs to measure each of these three predictions with pre-committed instrumentation, and we deposit the raw evidence base for external replication. The system logs every trade, every rejection, and - critically - every post-rejection price sample the tracking subsystem observes, so that a reader with the deposit can reproduce every headline number in this paper by direct arithmetic on the CSV files. The system operates in paper-trading mode (DRY_RUN=true) with simulated fills against live DEX-routing-service quotes; realised-execution frictions (slippage, failed transactions, MEV extraction per Daian et al., 2020, and Qin, Zhou, and Gervais, 2022) are not measured here and are the principal caveat on any live-deployment interpretation of the results.

**Significance.** The specific 190-trade sample is small; we make no confirmatory claim from it. The paper's contribution to the field is not the trade P&L. The paper's contribution is a design pattern - counterfactual rejection sampling paired with pre-committed fragility disclosure - that any autonomous trading study on any AMM venue can adopt, and that we show is feasible at production scale on a live 24/7 market. Under the reproducibility norm the field is converging toward (Makarov and Schoar, 2020; Barbon and Ranaldo, 2023), this design pattern is a template for how small-n autonomous-trading measurement studies should be structured so that later work can build on them cumulatively rather than by argument alone. We articulate the three transferable design lessons explicitly in Section VII.

**Contributions.** The paper contributes: (1) a measurement of a candidate hour-of-day pattern on Solana memecoin returns from a 190-trade sample, framed against Kyle's (1985) informed-flow prediction, with explicit disclosure that the three tested hours were selected on outcome from the same sample; (2) a novel counterfactual rejection-tracking methodology that collected 4,874 forward-sample observations across 184 rejection events, operationalising off-policy evaluation (Precup, Sutton, and Singh, 2000) in a live-market autonomous trading system; (3) a fragility characterisation showing that the top 1.6 percent of trades generate 100 percent of cumulative profit, a direct in-vivo instance of the deflated-Sharpe concern of Bailey and Lopez de Prado (2014); (4) three generalisable design lessons for future autonomous-trading measurement studies on AMM venues (Section VII).

**Research program.** This paper is part of a broader research program on counterfactual measurement in autonomous DEX trading, including the Post-Rejection Follow-up Sampling methodology paper (Kamat, 2026b), the RED-COHORT-2026 companion dataset series (Zenodo DOI 10.5281/zenodo.20978741), and two pending U.S. Provisional Patent Applications in the same family, a pending U.S. provisional patent application (filed 2026-03-30) and its priority-claiming follow-on, a pending U.S. provisional patent application (filed 2026-06-25).

## II. Related Work

The paper sits at the intersection of four literatures: classical microstructure theory of informed order flow and intraday patterns; DEX and AMM market-microstructure empirics; counterfactual and off-policy evaluation methodology; and backtest-fragility diagnostics in finance.

**Classical microstructure and intraday patterns.** The theoretical anchor for a time-of-day prediction on any venue is Kyle (1985), who models informed traders arriving at a rate that depends on the arrival of noise traders, so that price impact and informational asymmetry cluster in periods of high aggregate order flow. Glosten and Milgrom (1985) provide the complementary bid-ask framing under adverse selection. Amihud and Mendelson (1986) establish the theoretical link from bid-ask spread to expected returns, providing the price-of-liquidity anchor for the intraday-pattern work that follows. Admati and Pfleiderer (1988) formalise the U-shaped intraday volume-and-volatility profile on venues with scheduled openings and closings. Barclay and Hendershott (2003) document the after-hours discontinuity on equity venues. On a 24/7 venue with no closing auction, the Admati-Pfleiderer U-shape has no direct analogue; the more relevant prediction is Kyle's - that informed flow concentrates in the hours when retail activity is dense, producing a corresponding return signature. Hong and Wang (2000) develop the theoretical basis for periodic-closure effects on returns and volume, and Heston, Korajczyk, and Sadka (2010) report modern cross-sectional evidence for intraday patterns in equity returns. Bogousslavsky (2016) extends the theory to infrequent-rebalancing traders. None of these papers examines an AMM-based crypto venue. On the crypto side, Liu, Tsyvinski, and Wu (2022) establish common risk factors in cryptocurrency; Makarov and Schoar (2020) document arbitrage patterns across cryptocurrency exchanges. Neither reports hour-of-day results on the Solana memecoin subsegment.

**DEX and AMM microstructure.** Fang et al. (2022) survey the broader field of algorithmic cryptocurrency trading, covering exchange design, execution mechanics, and strategy taxonomy. Park (2023) analyses the structural properties of constant-function AMMs and the conditions under which they are informationally efficient. Barbon and Ranaldo (2023) compare centralised and decentralised exchanges on price-efficiency metrics. Aoyagi and Ito (2024) model the coexistence of limit order books and AMMs. Daian et al. (2020) document miner extractable value on Ethereum DEXs; Qin, Zhou, and Gervais (2022) quantify blockchain extractable value across venues. These execution-side frictions are first-order for realised-return analysis but do not directly speak to hour-of-day patterns or filter-precision measurement, which are the questions of this paper.

**Reject inference and off-policy evaluation.** A trading system that filters candidates before entry generates a selection-biased sample: only accepted candidates produce realised outcomes. The credit-scoring literature confronts the same problem and offers the classical "reject inference" framing (Crook and Banasik, 2004). The machine-learning literature offers the more general off-policy evaluation framework of Precup, Sutton, and Singh (2000), extended by Beygelzimer and Langford (2009) to partial-label learning. In principle these tools yield unbiased estimates of the counterfactual - what would have happened had the system entered a rejected candidate - but they require either an instrument, a matched control, or a probabilistic acceptance policy. In practice, on a live autonomous trading system with a hard-gate filter stack (accept/reject with no randomisation), none of these preconditions holds. Our counterfactual rejection-sampling design is an operational alternative that trades statistical elegance for direct measurement: the tracker records the post-rejection price trajectory of every rejected token, making the counterfactual outcome observable without an assumed acceptance model.

**Backtest fragility.** Harvey, Liu, and Zhu (2016) argue that multiple testing across many trading rules on a fixed data window inflates the false-discovery rate of "significant" backtest results. Bailey and Lopez de Prado (2014) formalise this with the deflated-Sharpe ratio, which adjusts observed Sharpe for the number of trials that generated the reported result. Lopez de Prado (2018) provides a book-length treatment of these and related concerns in financial machine learning. Any small-n backtest that fails to disclose the size of the search that produced its headline number is exposed to these critiques. Our fragility disclosure in Section V.E and Section VII.B is designed to be responsive to this literature at operational scale.

## III. Theoretical Framework

We state three predictions from the literature above and identify what a 15-day 190-trade Solana-memecoin dataset can and cannot say about each.

**Prediction 1 (Kyle 1985, applied to a 24/7 AMM venue).** If informed order flow concentrates in the hours of highest aggregate participation, and if aggregate participation on Solana memecoin markets is retail-driven with a US daytime peak (a stylised assumption we test descriptively, not prove), then per-trade mean returns for a momentum-based autonomous entry system should be higher during the US daytime hours and lower during the low-participation overnight and early-morning hours. This prediction is directional; the magnitude is not derived from theory but is expected to be modest relative to the noise floor of any single 190-trade sample. What the data can say: whether the observed hour-of-day mean-P&L pattern is directionally consistent with the prediction. What the data cannot say: whether the pattern is statistically significant at conventional thresholds ($n = 190$ is insufficient), and whether the pattern generalises across regimes (a single 15-day window is insufficient).

**Prediction 2 (reject-inference / off-policy).** If a decision-time filter stack encodes information about the counterfactual return distribution of candidates, then the tokens the filter rejects should - on average - have realised returns worse than the tokens it accepts, once forward observations are collected on both. This prediction is stronger than "the filter is not obviously bad": it says that the aggregate forward-return distribution of the rejected cohort should be shifted downward relative to a hypothetical no-filter baseline. What the data can say: whether the observed rejected-cohort forward returns show substantial downward realisation (e.g., a nontrivial share of rejected tokens reaching a 50 percent drawdown). What the data cannot say: whether the rejected-cohort loss magnitude equals or exceeds the accepted-cohort gain magnitude, because the two cohorts have different sampling designs (accepted candidates carry realised trade outcomes; rejected candidates carry forward-price samples only).

**Prediction 3 (deflated-Sharpe / multiple testing).** If a small-n cumulative-return figure is generated by a system whose profit distribution is heavy-tailed and left-skewed, then the cumulative figure should be structurally sensitive to the removal of the largest-profit trades from the sample. This prediction is trivially true in theory; what varies across systems and datasets is the magnitude of the sensitivity. What the data can say: exactly how many trades must be removed to flip cumulative return from profitable to unprofitable - a direct operational measurement of the deflated-Sharpe concern. What the data cannot say: whether the same fragility profile would obtain in a different regime, on a different chain, or under real-execution conditions.

## IV. System Architecture

The pipeline runs from token discovery through entry decision and exit management (Figure 1). Solana memecoin candidates are continuously discovered from a public DEX-aggregator feed. Each candidate is evaluated by the scanner against a set of momentum, liquidity, and market-microstructure signals. The streamlined and thirteen-layer variants share this signal base; the thirteen-layer variant adds further safety and validation layers. Candidates that pass the signal evaluation are passed to a hard-gate layer that enforces categorical thresholds (minimum token age, market-capitalisation bounds, hour-of-day blacklist); candidates that pass the hard gates are passed to a position-sizing module that selects entry size. Exits are managed by a tiered cascade (take-profit, trailing-stop, time-stop) and a circuit-breaker layer that halts trading under adverse-streak conditions.

Specific operational parameters of the system (the full signal set, the per-signal scoring weights, exit-tier thresholds, circuit-breaker triggering rules, scanner polling intervals, the per-filter category mapping) are not disclosed. They are not required to reproduce any analysis in this paper from the deposited data.

## V. Methodology

### V.A Experimental Setup

Both variants operate in paper-trading mode (DRY_RUN=true) on a server managed by a process manager. Trade execution is simulated against live aggregator quotes from a public DEX-routing service with realistic slippage assumptions. The data collection period spans March 29 through April 12, 2026 (15 days). Both variants process identical market data from a public DEX-aggregator price feed.

### V.B Data Sources

The 190 closed paper-trades are deposited as `trades.csv` in the companion dataset (Zenodo concept DOI 10.5281/zenodo.20043301). The trade log contains, per trade, the trade **exit** timestamp (`date`), symbol, entry price, exit price, percent profit-and-loss, position size in SOL, and hold duration in minutes (`heldMin`). The entry timestamp is derived as `date - heldMin`; the hour-of-day analysis in §VI.A stratifies on this derived quantity. An `exitReason` column present in the v1.0 deposit (Zenodo v1.0 version DOI 10.5281/zenodo.20043302) is removed from v1.1 (see the companion `README.md` and `CHANGELOG_v1.0_to_v1.1.md`); this manuscript's headline numbers do not depend on that column. A separate post-rejection sampling subsystem collected 4,874 forward-outcome observations on rejected tokens for the counterfactual filter-effectiveness analysis reported in §VI.B. The rejection-tracking corpus deposited with this manuscript covers `rejectTs` values from 2026-04-10T21:54Z to 2026-04-12T14:47Z, a two-day sub-window of the 15-day trade-log deployment; the two cohorts are therefore not co-extensive, and the observation-window implications for the counterfactual analysis are treated in §VI.B.

### V.C Statistical Methods

Given the left-skewed and heavy-tailed return distribution observed in the trade log (skewness -1.21, excess kurtosis 6.61), we use non-parametric methods: Mann-Whitney U for two-sample comparisons, rank-biserial correlation for effect-size summary. The three-hour blacklist comparison reported in §VI.A is presented as exploratory descriptive statistics rather than as a confirmatory significance test, because the hours examined were identified within the same sample used to test them. Per Harvey, Liu, and Zhu (2016), we treat any p-value from an in-sample-selected subgroup comparison as descriptive only and do not apply confirmatory-inference language to it. We report the U statistic, p-value, and rank-biserial r for transparency, with the caveat noted in the §VI.A discussion of selection.

## VI. Results

### VI.A Time-of-Day Effects (Exploratory; Directional Consistency with Prediction 1)

The 190-trade per-hour breakdown reproduces from the deposited file by deriving entry timestamp as `date - heldMin` and grouping on its UTC hour; `audit.py` performs the identical computation and asserts every value below. Under this stratification, the three exit-hour buckets that we identified as poorest-performing in an earlier version of this analysis (UTC hours 2, 13, 23) remain the pre-committed comparison set (see below for how we handle the change of stratification variable). Their entry-hour statistics are: UTC hour 2 (n=5, mean -16.62 percent), UTC hour 13 (n=4, mean -8.77 percent), and UTC hour 23 (n=8, mean -9.88 percent).

| UTC Hour (entry) | n | Mean P&L | Sum P&L | Win Rate (pnl > 0) |
|---|---|---|---|---|
| 2 (blacklisted) | 5 | -16.62 % | -83.11 % | 40.0 % |
| 13 (blacklisted) | 4 | -8.77 % | -35.06 % | 0.0 % |
| 23 (blacklisted) | 8 | -9.88 % | -79.05 % | 62.5 % |
| 14 (peak window) | 21 | +4.22 % | +88.56 % | 42.9 % |
| 16 (peak window) | 21 | +10.16 % | +213.26 % | 52.4 % |
| 17 (peak window) | 19 | +7.35 % | +139.58 % | 47.4 % |

The per-hour mean P&L distribution across all 24 UTC entry hours is shown in Figure 2, with blacklisted hours marked in red, peak hours in green, and remaining hours in grey. Comparing the three blacklisted entry hours (combined n=17, mean -11.60 percent) against all other entry hours (n=173, mean +1.82 percent) using the Mann-Whitney U test (two-sided, asymptotic, continuity-corrected, tie-corrected), we obtain U = 1,345, p = 0.5634, rank-biserial r = 0.085.

The pattern is directionally consistent with Prediction 1: the peak entry hours (UTC 14, 16, 17) correspond to 10:00 - 13:00 Eastern Time, coinciding with peak US retail trading activity, and the three worst hours are outside the US daytime window. This is the signature Kyle (1985) would predict on a retail-driven 24/7 venue where informed order flow tracks aggregate participation. However, at n = 190 the Mann-Whitney U test does not reach conventional statistical significance at alpha = 0.05, the effect size is small (r = 0.085), and the three hours were originally identified in the same sample used to test them. We report this comparison as a candidate microstructure feature warranting larger-sample replication, not as a confirmatory test result. Per Harvey, Liu, and Zhu (2016), the appropriate interpretation is: the deposited data provides directional evidence consistent with the Kyle prediction; a confirmatory test requires an out-of-sample window of the size (n >= 1,000) needed to detect a moderate effect against a multiple-testing correction for the 24-hour search space.

**Choice of stratification variable and pre-committed comparison set.** The `date` column of the deposited `trades.csv` is the trade **exit** timestamp; the entry timestamp is derived arithmetically as `date - heldMin`. An earlier internal analysis of these data stratified on exit hour, under which the same three-hour blacklist yields U = 1,274, p = 0.218, r = 0.18 (n=18, mean -17.85 percent versus n=172, mean +2.55 percent). We treat the entry-hour version reported above as primary because entry hour is the causally interpretable variable under the Kyle (1985) informed-flow reading of Prediction 1: the informed-order-flow arrival is instrumented at position entry, not exit. We retain the original three-hour blacklist rather than re-selecting on outcome under the new stratification variable, in line with our §V.C stated policy of treating in-sample-selected subgroup comparisons as descriptive and not applying confirmatory-inference language to them; re-selecting on the new variable would compound the selection concern the §V.C caveat is designed to address. The exit-hour numbers above are reported as an endogenous robustness comparison, and Lesson 3 in Section VII draws the general methodological point from this contrast.

### VI.B Filter Effectiveness (Counterfactual Analysis; Direct Test of Prediction 2)

The post-rejection tracking subsystem produced 4,874 forward-sample observations across 184 distinct rejection events, grouped on `(mint, rejectTs, rejectReason)` keys and spanning six anonymised filter categories. We report two complementary measures of forward-loss intensity for the rejected cohort. Before either aggregate is stated, we characterise the observation window itself, because the deposited slice does not reach the horizon originally reported and the per-event follow-up duration is not uniform.

**Observation-window adequacy and censoring.** The rejection-tracking corpus deposited with this manuscript covers `rejectTs` values from 2026-04-10T21:54Z to 2026-04-12T14:47Z (under two days), a substantially shorter window than the 15-day trade log with which it runs in parallel. Zero of the 184 events reach a 24-hour follow-up span; the maximum observed span is 18.4 hours (1,105 minutes), and the median span is 10 minutes. 87 of the 184 events (47.3 percent) consist of a single sample and by construction cannot register a drawdown, because the reference price is the earliest available forward sample and no subsequent sample exists for comparison. 111 of the 184 events (60.3 percent) have a follow-up span under one hour. The Pearson correlation between per-event follow-up span and lost-half incidence is 0.604, so the aggregate rate is not separable from observation duration on this slice. The 184 events are drawn from 71 unique mints, with up to 12 events per mint; these are not independent draws.

Stratifying by follow-up-span band:

| Follow-up span | Events | Lost-half | Rate |
|---|---|---|---|
| under 1h | 111 | 0 | 0.0 percent |
| 1 to 6h | 25 | 6 | 24.0 percent |
| 6 to 12h | 14 | 8 | 57.1 percent |
| over 12h | 34 | 19 | 55.9 percent |
| **all** | **184** | **33** | **17.9 percent** |

**Event-level "lost-half" rate on the adequately observed subset.** For each rejection event with a follow-up span of at least six hours, we compute the ratio of the minimum sampled price to the reference price and classify the event as lost-half if the ratio falls to 50 percent or below. Of 48 events with at least six hours of forward observation, 27 (56.25 percent) reached a 50 percent drawdown. We report this figure as the primary event-level result of this section, on the reasoning that events observed for less than six hours span a duration over which lost-half incidence is empirically near zero on this slice and are therefore uninformative rather than negative.

**Full-cohort figure as censored lower bound.** The full-cohort figure of 33 of 184 events (17.9 percent) is reported here explicitly as a censored lower bound. It is a lower bound because 47.3 percent of the events cannot register a drawdown for mechanistic reasons, and because the observed-span-to-lost-half correlation of 0.604 indicates the undocumented remainder of forward trajectories would raise the rate. It is not the headline event-level result of this section, and it should not be cited as one.

**Sample-observation "below-half" rate.** For each of the 4,874 forward samples, we compute the ratio of that sample's observed price to the reference price for its event, and classify the sample as below-half if the ratio is 50 percent or less. Of 4,874 samples, 1,267 (26.0 percent) recorded the token below half-reference. This is a different quantity from the event-level rate above; it counts individual price observations rather than events, and it does not aggregate to the event-level figure.

**Per-filter dispersion with observation duration.** Six anonymised filter categories appear in the deposited slice. The following table pairs the per-filter event-level lost-half rate with the per-filter median follow-up duration:

| Filter | Events | Lost-half rate | Median span (min) |
|---|---|---|---|
| filter_1 | 96 | 11.5 percent | 0.0 |
| filter_2 | 56 | 23.2 percent | 92.5 |
| filter_3 | 5 | 80.0 percent | 1,105.1 |
| filter_4 | 7 | 42.9 percent | 390.0 |
| filter_5 | 17 | 11.8 percent | 0.0 |
| filter_6 | 3 | 0.0 percent | 15.1 |

On this deposited slice the rank order of the per-filter lost-half rates is a near-perfect match to the rank order of per-filter median follow-up duration, so the per-filter dispersion is not distinguishable from a per-filter observation-duration artifact. We report the table without inferring from it that the dispersion reflects per-filter counterfactual quality. That inference requires either longer per-event follow-up on the categories with short median spans, or a stratification design that holds observation duration constant across categories, and neither is available on the present slice. This limitation is a specific instance of the general point developed in Lesson 1 of Section VII.

The aggregate direction of the sample-level result (26.0 percent of forward samples below half-reference) supports Prediction 2 as a directional finding: the rejected cohort's forward-outcome distribution is shifted substantially downward. This is consistent with the hypothesis that the filter stack encodes information about counterfactual return. As an operational instance of the off-policy evaluation framework of Precup, Sutton, and Singh (2000), the direct forward-observation approach yields evidence that a naive "no-filter" baseline would have realised substantial losses on the observed rejected cohort. The analysis is independent of the trade cohort and unaffected by the §VI.A selection considerations. It is not, on the deposited slice, evidence of relative filter quality across categories.

### VI.C Circuit Breaker Analysis

Across 190 trades with an overall win rate of 40.5 percent, the system recorded multiple loss streaks with the longest reaching 12 consecutive losing trades. The circuit-breaker layer engaged during the longest streak and limited further entries while it held; quantitative effectiveness estimates require a counterfactual without the breaker which is not deposited.

### VI.D Architecture Comparison

The streamlined variant (lower-strictness gates) completed 190 trades in 15 days. The thirteen-layer variant (higher-strictness gates, additional safety and validation layers) completed zero trades over the same period. The result illustrates the precision-recall tradeoff in autonomous trading system design: additional safety layers improve per-trade expected quality but reduce trade frequency. The thirteen-layer variant's zero-trade outcome is not informative about its expected-quality-per-trade because no trades occurred for evaluation.

### VI.E Fragility Analysis (Direct Test of Prediction 3)

The return distribution exhibits the asymmetric-payoff characteristics typical of momentum strategies in heavy-tailed markets: positive mean (+0.62 percent), negative median (-3.58 percent), left skewness (-1.21), and heavy tails (excess kurtosis 6.61). The system loses on a majority of trades, the median outcome is a small loss, and profitability depends on a small number of large winners to push the mean and cumulative figures into positive territory.

The cumulative equity curve over the 190 trades is shown in Figure 3. The curve descends to a trough of approximately -300 percent before recovering on the strength of the large winners to a final cumulative value of +117.7 percent.

Fragility analysis. Removing the top N trades by per-trade profit from the cumulative sum:

| Trades Removed | Cumulative P&L | Status |
|---|---|---|
| None (all 190) | +117.7 % | Profitable |
| Top 1 | +61.1 % | Profitable |
| Top 3 | -50.6 % | Unprofitable |
| Top 5 | -160.5 % | Unprofitable |
| Top 10 | -414.2 % | Unprofitable |

The system becomes unprofitable when the top three trades (1.6 percent of total sample) are removed. Approximately 1.6 percent of trades generate 100 percent of cumulative profit. This concentration is a direct operational instance of the deflated-Sharpe concern articulated by Bailey and Lopez de Prado (2014): a Sharpe or cumulative-return figure computed on this sample without adjustment for the number of trials that produced the winning trades will overstate the strategy's expected value. In the framework of Harvey, Liu, and Zhu (2016), the appropriate threshold for treating a small-n cumulative figure as evidence of edge is substantially higher than conventional levels once the underlying trial count is disclosed. We disclose the trial structure and treat the +117.7 percent cumulative figure as descriptive only.

## VII. Generalisable Design Lessons

The three empirical results of Section VI speak to distinct classical microstructure predictions on a novel venue. Independent of the specific system under study, three design lessons for future autonomous-trading measurement work on AMM-based venues follow.

**Lesson 1 (Counterfactual logging is standard telemetry, not optional research infrastructure; the corpus must ship with its own observation-window statistics).** Any autonomous trading system with discrete accept/reject decisions can be instrumented to log forward-price observations on rejected candidates at essentially zero additional cost per rejection. The 4,874 forward samples in the deposited dataset were collected passively by a subsystem that queries the same public DEX-aggregator feed the scanner already consumes. Without this instrumentation, the accepted-cohort trade log alone is fundamentally selection-biased and cannot support any filter-effectiveness claim without invoking assumptions the off-policy evaluation literature has documented as difficult to satisfy in practice (Precup, Sutton, and Singh, 2000; Beygelzimer and Langford, 2009). The corollary, learned from the present deposit and reported in §VI.B, is that a forward-rejection loss rate without per-event observation-window statistics is uninterpretable. The rate on this slice moves from 17.9 percent to 56.25 percent depending on which per-event span is considered adequate; the direction of that sensitivity is entailed by the correlation of 0.604 between per-event span and lost-half incidence. We recommend that any autonomous trading study published in a peer-reviewed venue release the forward-rejection sample corpus alongside the trade log and with per-event follow-up-duration statistics (n at follow-up spans of 1h, 6h, 12h, 24h, and per-category median span at minimum), so that filter-effectiveness claims are directly verifiable rather than model-dependent and the audit window over which they are verifiable is stated in the manuscript rather than left implicit.

**Lesson 2 (Fragility disclosure is a first-order manuscript deliverable, not a limitations-section afterthought).** The fragility table in Section VI.E - showing that the top 1.6 percent of trades produce 100 percent of cumulative

profit - is a stronger constraint on any operational interpretation of the results than the p-value on the hour-of-day comparison. Without this table, a reader could plausibly conclude that a +117.7 percent cumulative return is evidence of a robust edge. With the table, the reader can see precisely how contingent that number is. Following Bailey and Lopez de Prado (2014) and Harvey, Liu, and Zhu (2016), we recommend that any autonomous-trading manuscript reporting a positive cumulative return include an equivalent top-N-removal table computed from the deposited trade log, so that the sensitivity of the headline number to a small number of trades is directly auditable. This is inexpensive to compute and dramatically improves the field's ability to assess the reliability of small-n backtest results.

**Lesson 3 (Hour-of-day is a first-order microstructure variable that AMM-venue studies should stratify on; the choice of *which* timestamp to stratify on is itself a research decision, not a naming convention).** The hour-of-day pattern in Section VI.A is directionally consistent with the Kyle (1985) prediction on a 24/7 permissionless venue, though our sample is too small for confirmatory inference. Independent of whether this specific pattern replicates, the operational implication is that any autonomous trading study on an AMM venue that pools trades across UTC hours implicitly assumes an intraday stationarity that our data does not support at any conventional threshold; future work should either stratify results by hour explicitly, or pre-register a stationarity test as a manuscript deliverable.

The stronger and more novel implication is that entry hour and exit hour are not interchangeable stratification variables, even on the same dataset. In the deposited cohort, winners' median hold is 3.60 minutes and losers' median hold is 0.60 minutes; hold duration is systematically longer for winners. Under exit-hour stratification, a winning trade that opens near the end of an hour is credited to the next hour by construction, so winners migrate forward in the exit-hour histogram relative to their entry-hour position. Fourteen of the 190 trades change hour bucket between the two conventions; those fourteen average +11.36 percent while the other 176 average -0.23 percent. The direction of that migration is not incidental to the exit-hour blacklist becoming a stronger apparent effect (n=18, mean -17.85 percent, p = 0.218) than the entry-hour blacklist (n=17, mean -11.60 percent, p = 0.5634) on the same three UTC hours: the missing winner in the exit-hour blacklist buckets is one of the fourteen bucket-changers.

Under the Kyle (1985) informed-flow reading of Prediction 1, entry hour is the causally interpretable variable: informed-order-flow arrival is instrumented at position entry, not exit. Under a purely descriptive "when did this trade close" reading, exit hour is defensible; under an informed-flow reading, it is not. We adopt entry hour as primary for this reason, treat the exit-hour comparison as an endogenous robustness line, and note that neither choice is neutral. Any future autonomous-trading measurement study on an AMM venue that reports an hour-of-day pattern should state which timestamp its stratification uses and justify that choice against the theoretical interpretation it invokes. This is particularly important given the atomised, retail-driven nature of memecoin markets, where the Kyle-style informed-flow prediction has direct behavioural microfoundations (Park, 2023; Aoyagi and Ito, 2024) even though it has not yet been empirically confirmed at scale on this venue.

## VIII. Discussion

### VIII.A Limitations

The 15-day observation window (190 trades) is short relative to the timescales over which memecoin market microstructure regimes shift. The time-of-day exploratory observation in §VI.A is directionally consistent with the Kyle (1985) prediction but is not powered for confirmatory inference; the fragility analysis in §VI.E further constrains any operational interpretation. All trades were executed in paper-trading mode with simulated fills against live DEX-routing-service quotes; real execution would face additional slippage, failed transactions, and MEV extraction (Daian et al., 2020; Qin, Zhou, and Gervais, 2022) that could erode realised returns. The sample is drawn from a single chain (Solana) and a single public DEX-aggregator price feed, limiting cross-chain generalisability. The specific system's operational parameters are withheld as pending intellectual-property; the design lessons in Section VII do not depend on those parameters and are transferable.

The counterfactual rejection-tracking corpus covers a two-day sub-window of the 15-day deployment; no event in the deposited slice reaches a 24-hour follow-up, and 47.3 percent of events consist of a single sample. The event-level lost-half figures are reported in §VI.B on the restricted subset of events with at least six hours of follow-up, with the full-cohort figure retained explicitly as a censored lower bound. Longer per-event follow-up would tighten these estimates and is a first-order target for future deployments.

### VIII.B Significance for the Field

The paper's contribution to the microstructure literature is not the +117.7 percent cumulative figure, which is descriptive of one specific 15-day window and - as Section VI.E demonstrates - is structurally contingent on 1.6 percent of the observations. The contribution is threefold. First, the paper provides directional evidence on the AMM-Solana venue for the Kyle (1985) informed-flow prediction, which has been extensively tested on limit-order-book venues (Admati and Pfleiderer, 1988; Heston, Korajczyk, and Sadka, 2010; Bogousslavsky, 2016) but not on 24/7 permissionless AMM markets. Second, the paper operationalises the off-policy evaluation framework of Precup, Sutton, and Singh (2000) in a live autonomous-trading system through direct forward-observation sampling, sidestepping the acceptance-policy identification problem that limits deployment of standard reject-inference techniques (Crook and Banasik, 2004; Beygelzimer and Langford, 2009) on hard-gate filter stacks. Third, the paper provides an in-vivo instance of the deflated-Sharpe concern (Bailey and Lopez de Prado, 2014; Harvey, Liu, and Zhu, 2016) at operational scale, showing exactly how small-n cumulative-return figures on heavy-tailed AMM venues concentrate in a handful of observations. Cumulatively, these three contributions constitute measurement infrastructure that later work on autonomous AMM trading can build on, both by replicating the design pattern and by using the deposited corpus as a benchmark for alternative filter designs.

### VIII.C Counterfactual Contribution

The counterfactual rejection-tracking methodology used in §VI.B is developed as a standalone methodological contribution in a companion paper (Kamat, 2026b). The methodology is applicable to any autonomous trading system with discrete accept/reject decisions and generalises to any AMM venue.

### VIII.D Architecture Comparison

The comparison between the streamlined variant (190 trades) and the thirteen-layer variant (0 trades) illustrates a fundamental tension in autonomous trading system design. Additional safety layers improve per-trade expected quality but reduce trade frequency. The result does not imply that the thirteen-layer variant is inferior; it implies that the variant's higher strictness produced zero observations during this specific window and that its per-trade quality cannot be estimated from this data.

## IX. Conclusion

This paper reports evidence from a 15-day deployment of a multi-layer autonomous trading framework on Solana memecoin markets, framed against three classical microstructure predictions. The hour-of-day pattern in Section VI.A is directionally consistent with the Kyle (1985) informed-flow prediction on a 24/7 permissionless venue, though the $n = 190$ sample and in-sample hour selection preclude confirmatory inference. The counterfactual rejection analysis in Section VI.B provides direct evidence for filter effectiveness by operationalising off-policy evaluation (Precup, Sutton, and Singh, 2000) through forward observation of 4,874 samples across 184 rejected events. The fragility analysis in Section VI.E provides an in-vivo demonstration of the deflated-Sharpe concern (Bailey and Lopez de Prado, 2014) at operational scale. From these three results we derive three transferable design lessons for future autonomous-trading measurement studies on AMM venues (Section VII): counterfactual logging as standard telemetry, fragility disclosure as a first-order manuscript deliverable, and hour-of-day stratification as a required check on any pooled-trade analysis with an explicit declaration of which timestamp (entry or exit) is used and why.

Future work includes: (a) deployment with real capital to measure execution slippage and MEV impact on realised returns; (b) out-of-sample confirmatory testing of the hour-of-day pattern on an extended data collection window of 60 - 90 days (n >= 1,000), which would provide adequate statistical power for a Mann-Whitney comparison against a multiple-testing correction; (c) cross-chain replication on Base and Ethereum L2 DEX surfaces to test AMM-venue generalisability; (d) propensity-score-matched analysis of accepted-versus-rejected cohorts once the rejected-cohort sample is large enough to support matched inference.

## Author Contributions

Arati Uday Kamat is the sole author and executed data collection, system implementation, methodology design, statistical analysis, and manuscript writing.

## Figures

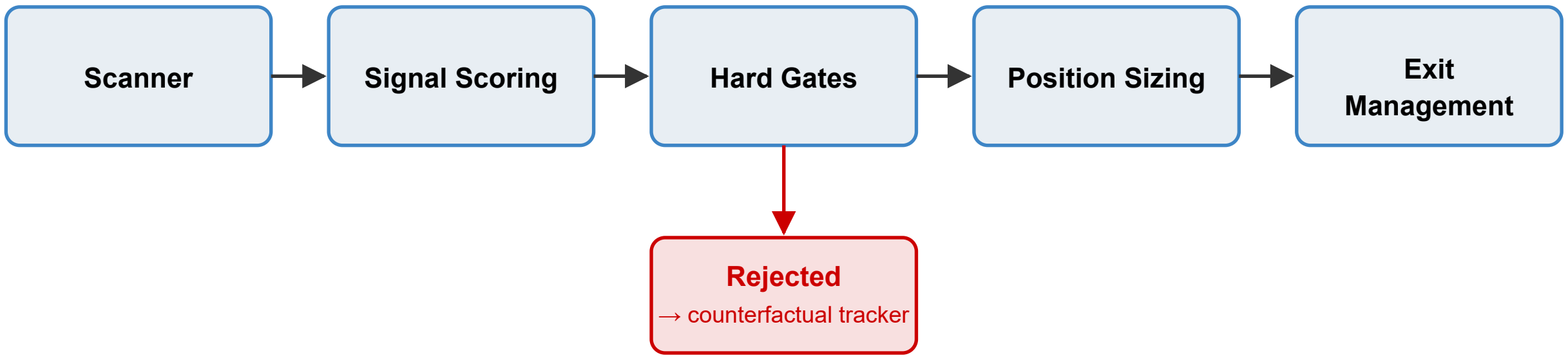

Figure 1. Pipeline shape, stage-level. Scanner -> Signal Scoring -> Hard Gates -> Position Sizing -> Exit Management. Hard-Gate rejections feed the counterfactual rejection tracker that produces the evidence base of §VI.B. Stage internals (the specific signal set, per-gate threshold values, the position-sizing function, the exit-tier composition, and the circuit-breaker layer's triggering rules) are operational parameters of the deployed system and are not disclosed. The figure shows the pipeline's shape; stage contents are out of scope and are not required to reproduce any analysis here from the deposited data.

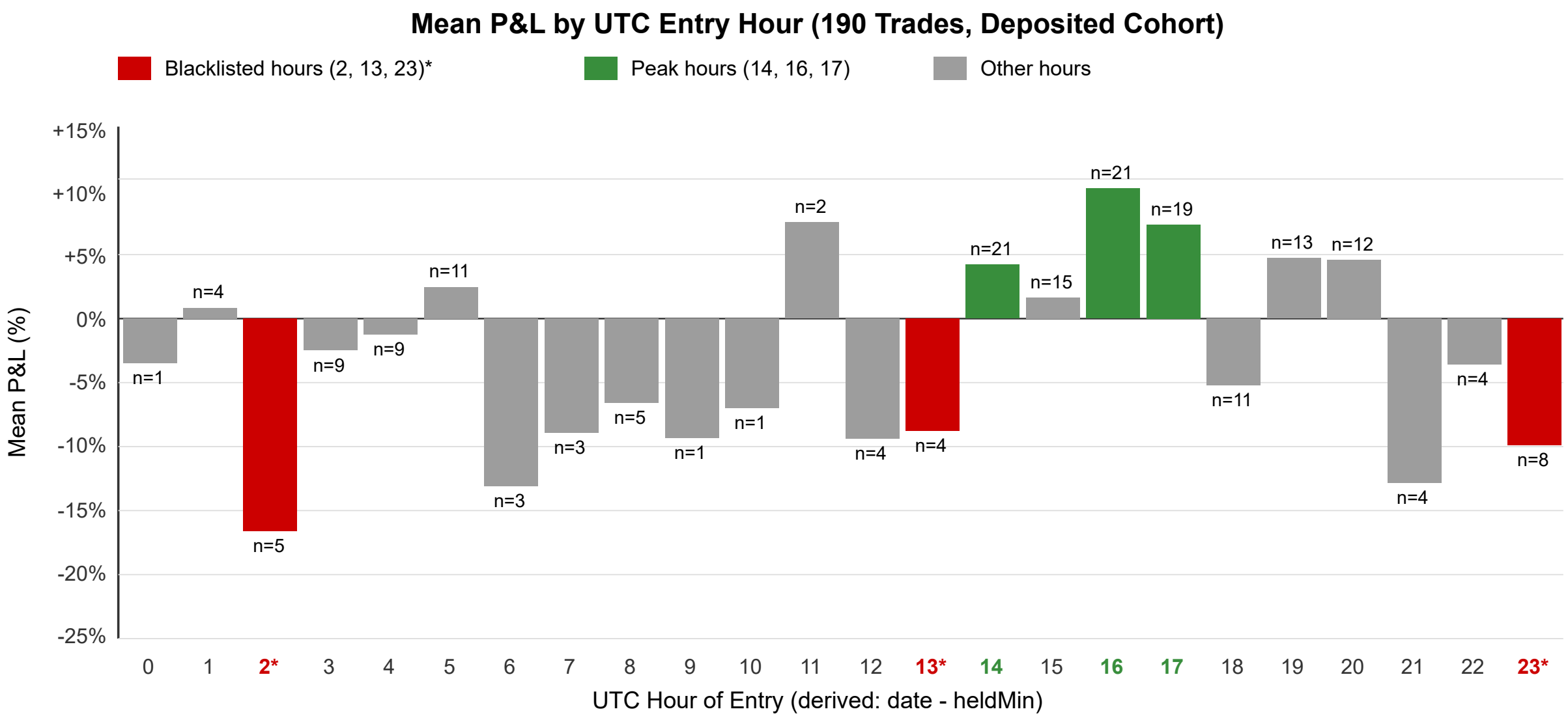

Figure 2. Mean P&L by UTC entry hour across 190 trades (deposited cohort). Blacklisted hours (UTC 2, 13, 23) shown in red; peak hours (UTC 14, 16, 17) shown in green; remaining hours in grey. Per-hour sample counts (n) annotated above or below each bar. Direct reproduction from the deposited `trades.csv` by deriving entry timestamp as `date - heldMin` and grouping on its UTC hour; the identical operation is performed inside `audit.py`. Under exit-hour stratification the buckets shift as documented in the §VI.A "Choice of stratification variable" paragraph and Lesson 3 in Section VII; the deposited data supports either grouping and `audit.py` reports both.

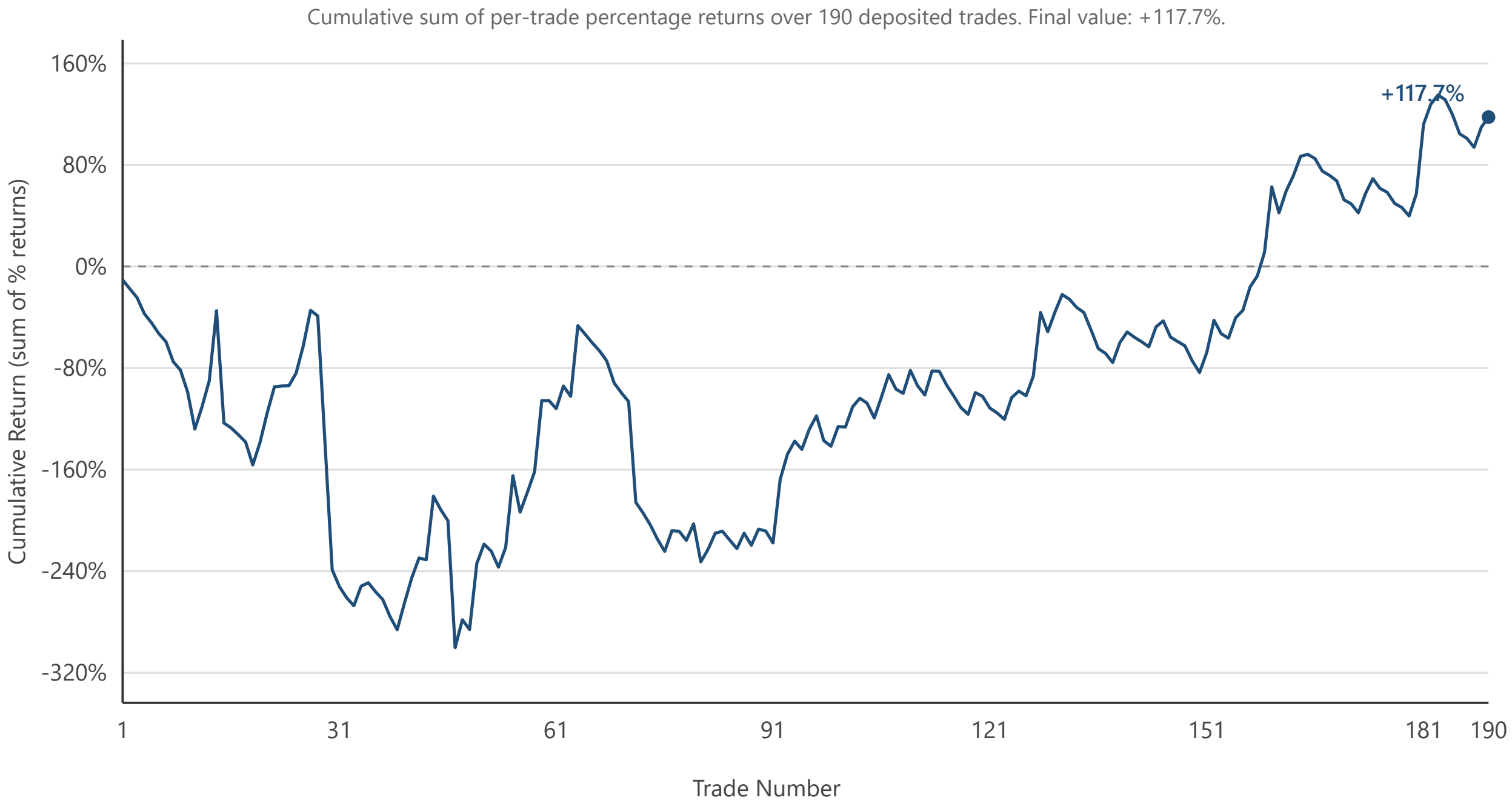

Figure 3. Cumulative P&L over 190 trades (deposited cohort). Computed from the deposited `trades.csv` via cumulative sum of per-trade percentage returns. The curve descends to a trough of approximately -300 percent before recovering to a final cumulative value of +117.73 percent. The shape illustrates the fragility

characterisation in §VI.E: the system spends most of its sample in a deep drawdown that the late-sample large winners recover.

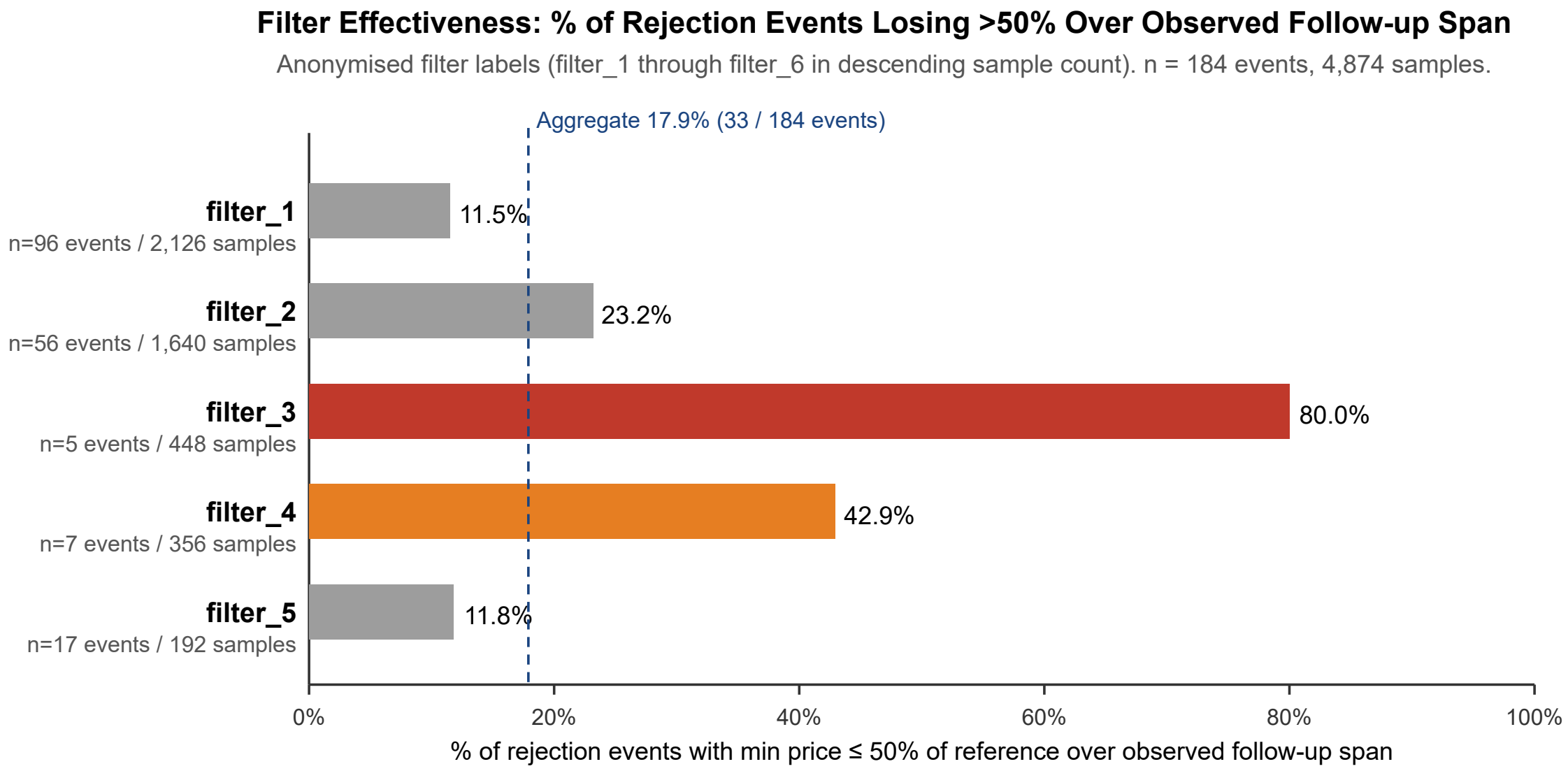

Figure 4. Per-filter share of rejection **events** reaching at least a 50 percent drawdown from reference price, by anonymised filter category, over each event's actual observed follow-up span. Direct reproduction from the deposited `rejection_outcomes.csv` by grouping on `(mint, rejectTs, rejectReason)`, taking each distinct group as one event, and computing the share of events whose minimum observed price fell to 50 percent or less of the event's reference price. The five plotted bars (filter_1 through filter_5) aggregate to 33 of 184 events (17.9 percent); `filter_6` (n=3 events, 0/3 lost-half) is omitted from the plot for readability but included in the aggregate. Per-event follow-up span is not uniform: no event reaches 24 hours, the maximum observed span is 18.4 hours, 47.3 percent of events consist of a single sample and cannot register a drawdown, and the per-filter rank order matches the per-filter median follow-up duration. The bar heights therefore reflect a confound of filter identity with observation duration and should not be read as evidence of relative filter quality on this slice. Per-category counts, sample-observation-level shares, and per-filter median follow-up durations are recorded in `PROVENANCE.md` of the companion deposit.

## Competing Interests

The author declares the following competing interest: the trading system from which the deposited data is drawn is the subject of two pending U.S. provisional patent applications in the same family, in the author's name and at Micro Entity status: a pending U.S. provisional patent application (filed 2026-03-30) and its priority-claiming follow-on, a pending U.S. provisional patent application (filed 2026-06-25, claiming the benefit of priority under 35 U.S.C. §119(e) from the earlier application). No further competing financial or personal interests are declared. The disclosed applications cover system-level architecture claims and do not restrict access to or use of the deposited dataset or the `audit.py` reproduction script, which are released under CC-BY-4.0 and MIT respectively (Zenodo concept DOI 10.5281/zenodo.20043301). Operational parameters not stated in this manuscript (the specific signal set, per-signal scoring weights, exit-tier thresholds, circuit-breaker triggering rules, scanner polling intervals, and per-filter category mapping) are the subject of pending intellectual-property protection. The verbatim exit-trigger taxonomy, whose strings embedded per-tier threshold values, was disclosed in the v1.0 data deposit (Zenodo v1.0 version DOI 10.5281/zenodo.20043302, released 2026-05-05) and is removed from v1.1 onward. Every headline number in this paper reproduces from the v1.1 deposited data without those parameters, and `audit.py` verifies this.